\documentclass[aip,apl,amsmath,amssymb,reprint,showkeys]{revtex4-1}

\usepackage{graphicx}
\usepackage{dcolumn}
\usepackage{bm}

\begin{document}


\title{The quasi-free-standing nature of graphene on H-saturated SiC(0001)}

\author{F. Speck}
\affiliation{Lehrstuhl f\"ur Technische Physik, Universit\"at Erlangen-N\"urnberg, Erwin-Rommel-Str. 1, 91058 Erlangen, Germany}
\author{J. Jobst}
\affiliation{Lehrstuhl f\"ur Angewandte Physik, Universit\"at Erlangen-N\"urnberg, Staudtstr. 7, 91058 Erlangen, Germany}
\author{F. Fromm}
\affiliation{Lehrstuhl f\"ur Technische Physik, Universit\"at Erlangen-N\"urnberg, Erwin-Rommel-Str. 1, 91058 Erlangen, Germany}
\author{M. Ostler}
\affiliation{Lehrstuhl f\"ur Technische Physik, Universit\"at Erlangen-N\"urnberg, Erwin-Rommel-Str. 1, 91058 Erlangen, Germany}
\author{D. Waldmann}
\affiliation{Lehrstuhl f\"ur Angewandte Physik, Universit\"at Erlangen-N\"urnberg, Staudtstr. 7, 91058 Erlangen, Germany}
\author{M. Hundhausen}
\affiliation{Lehrstuhl f\"ur Technische Physik, Universit\"at Erlangen-N\"urnberg, Erwin-Rommel-Str. 1, 91058 Erlangen, Germany}
\author{H.~B. Weber}
\affiliation{Lehrstuhl f\"ur Angewandte Physik, Universit\"at Erlangen-N\"urnberg, Staudtstr. 7, 91058 Erlangen, Germany}
\author{Th. Seyller}
\affiliation{Lehrstuhl f\"ur Technische Physik, Universit\"at Erlangen-N\"urnberg, Erwin-Rommel-Str. 1, 91058 Erlangen, Germany}
\email{thomas.seyller@physik.uni-erlangen.de}

\date{\today}

\def\DEGC{$^\circ$C}
\def\OUT{$(6\sqrt3\times6\sqrt3)R30^{\circ}$}
\def\6R3{$6\sqrt3$}
\def\R3{$(\sqrt3\times\sqrt3)R30^{\circ}$}
\def\3BY3{$(3\times3)$}
\def\CFACE{$(000\overline{1})$}
\def\SIG{$\sigma$}
\def\PI{$\pi$}
\def\PS{$\pi^*$}
\def\EF{$E_{\mathrm{F}}$}
\def\ED{$E_{\mathrm{D}}$}
\def\kpar{$k_{\parallel}$}
\def\kperp{$k_{\perp}$}
\def\ZERO{'0$^{th}$'}
\def\MOB{cm$^2$/Vs}
\def\CUTOFF{{$f_{T}$}}

\begin{abstract}

We report on an investigation of quasi-free-standing graphene on 6H-SiC(0001) which was prepared by intercalation of hydrogen under the buffer layer. Using infrared absorption spectroscopy we prove that the SiC(0001) surface is saturated with hydrogen. Raman spectra demonstrate the conversion of the buffer layer into graphene which exhibits a slight tensile strain and short range defects. The layers are hole doped ($p = 5.0-6.5 \times 10^{12}$~cm$^{-2}$) with a carrier mobility of 3,100~{\MOB} at room temperature. Compared to graphene on the buffer layer a strongly reduced temperature dependence of the mobility is observed for graphene on H-terminated SiC(0001) which justifies the term "quasi-free-standing".
\end{abstract}


\maketitle

\begin{figure}
\includegraphics[width=6.0cm]{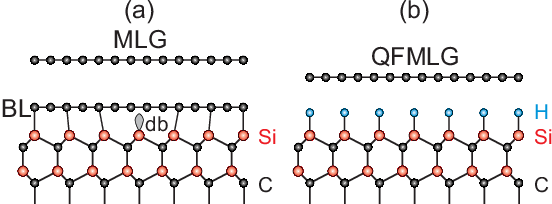}
\caption{\label{fig:structure}Structure of (a) monolayer graphene (MLG) on the buffer layer (BL) and (b) quasi-free-standing graphene (QFMLG) on hydrogen saturated SiC(0001).}
\end{figure}

Epitaxial graphene (EG) on SiC surfaces\cite{berger2006a,emtsev2009a,first2010a} paves the way for technological applications of graphene in, e.g., high frequency transistors \cite{lin2010a} or resistance standards.\cite{tzalenchuk2010a} Consequently, EG has been studied in various aspects (see, e.g., ref. \cite{first2010a} and references therein) and it was shown that there are significant differences in the material's properties depending on whether graphene is grown on the Si-terminated or the C-terminated surface of the SiC substrate. We focus on the Si-face, where graphene monolayers (called MLG in the following) can be grown routinely. In this material an electrically insulating buffer layer (BL) with {\OUT} periodicity exists at the interface. One can think of the BL as a graphene layer covalently bound to the top Si atoms of the substrate \cite{emtsev2008a} as is shown schematically in fig. \ref{fig:structure}(a). Compared with ideal graphene, MLG shares the same band structure,\cite{bostwick2007a} Raman signature,\cite{roehrl2008a} and graphene-specific quantum Hall effect.\cite{jobst2010a,tzalenchuk2010a,shen2009a} However, the substrate induces certain perturbations. MLG is strongly electron doped \cite{bostwick2007a}  with a concentration of $n \approx 1 \times 10^{13}$~cm$^{-2}$. A model was proposed\cite{kopylov2010a} which  describes the charge transfer from the substrate to MLG by assuming donation of electrons either from bulk donors or from states at the interface. Because of the lattice mismatch between the BL and the SiC(0001) surface it is reasonable to assume a sufficiently high density of amphoteric dangling bonds (db) in the interface region to provide the Fermi-level pinning observed in MLG. 

The intrinsic carrier concentration in MLG affects the carrier mobility $\mu$. Pristine MLG with $n \approx 1 \times 10^{13}$~cm$^{-2}$ has a mobility of around 2,000~{\MOB} at $T = 25$~K.\cite{emtsev2009a,jobst2010a} Doping MLG with an overlayer of tetrafluoro-tetracyanoquinodimethane to $n \approx 1 \times 10^{11}$~cm$^{-2}$ yields a carrier mobility of $\mu = 29,000$~{\MOB} at $T=25$~K \cite{jobst2010a} which is similar to what is observed for exfoliated graphene on SiO$_2$ substrates close to charge neutrality.\cite{novoselov2005a,zhang2005a} What sets MLG on SiC(0001) aside from exfoliated graphene is its strong temperature dependence of $\mu$. \cite{emtsev2009a,jobst2010a,weingart2010a} For pristine MLG with $n \approx 1 \times 10^{13}$~cm$^{-2}$ the mobility drops to values of around 900~{\MOB} at room temperature (RT), even when atomically flat Hall bars are prepared.\cite{jobst2010a} The $T$ dependence of the carrier mean free path indicates strong electron-phonon scattering presumably involving substrate phonons. It is thus desirable to reduce the coupling to the substrate while keeping the structural quality and the epitaxial character intact.

Riedl et al.\cite{riedl2010a} intercalated hydrogen underneath the BL, which subsequently converts into graphene.  They baptized the material quasi-free-standing monolayer graphene (QFMLG). It sits on a H-terminated SiC(0001) surface as depicted in fig. \ref{fig:structure}(b). The presence of H was concluded from comparison of the Si 2p core level spectrum with that of H-terminated SiC(0001).\cite{sieber2003a} They observed that the charge density of QFMLG was strongly reduced, in some cases even reversed in sign (hole doping) and suggested that this method is a viable route for tailoring the interface between SiC and graphene.\cite{riedl2010a} At this point, further analysis is required to judge how far QFMLG can be considered as quasi-free-standing.

\begin{figure}
\includegraphics[width=6.0cm]{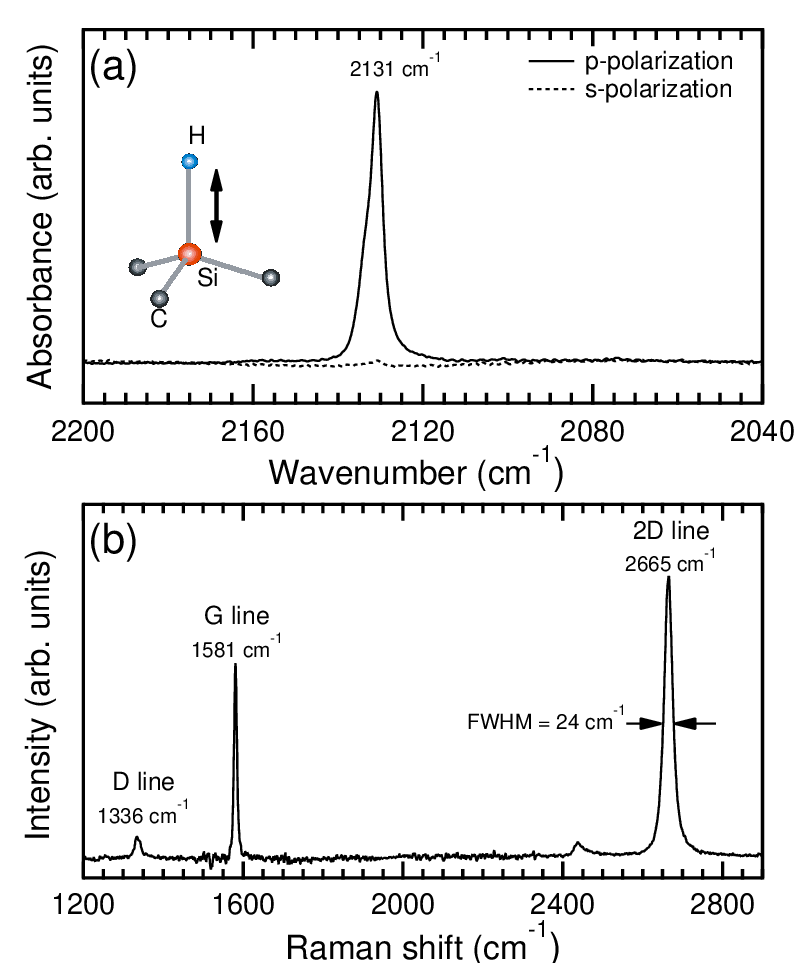}
\caption{\label{fig:spectrum}(a) FTIR-ATR spectrum of QFMLG showing the Si-H stretch mode. (b) Typical Raman spectrum of QFMLG.}
\end{figure}

\begin{figure}
\includegraphics[width=6.0cm]{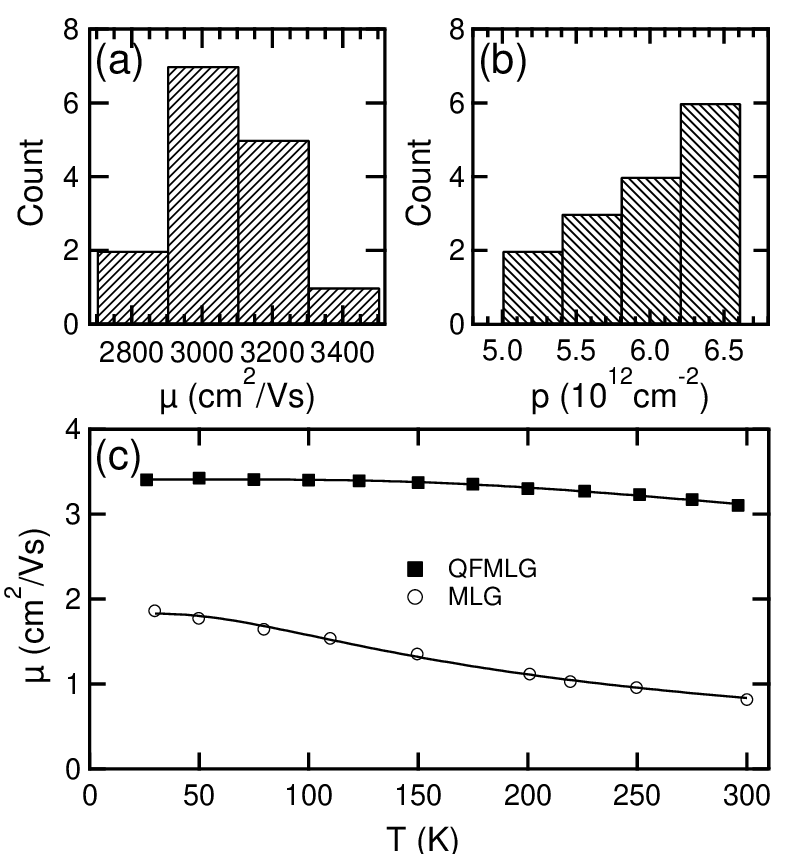}
\caption{\label{fig:transport}Transport data of QFMLG. Histograms of (a) carrier mobility $\mu$ and (b) hole density $p$ of 15 Hall bars made from QFMLG. (c) Temperature dependence of $\mu$ of QFMLG with $p=5.7\times 10^{12}$~cm$^{-2}$ compared to MLG [ref. 9] with $n=1\times 10^{13}$~cm$^{-2}$. Solid lines are fits according to ref. 28.}
\end{figure}

Here we examine the nature of QFMLG using infrared absorption spectroscopy in the attenuated total reflection mode (FTIR-ATR), Raman spectroscopy, and Hall effect measurements. FTIR-ATR spectroscopy was carried out at RT with the sample mounted face down on a germanium prism. Raman spectra were measured also at RT using a laser wavelength of 532~nm. The carrier concentration and mobility was determined by Hall effect measurements in the temperature range between 25 K and 300 K. The preparation of Hall bars by e-beam lithography is described in refs. \cite{emtsev2009a,jobst2010a}. The sample preparation included the growth of the BL followed by intercalation of hydrogen. The BL was prepared by annealing on-axis oriented 6H-SiC(0001) samples in 1000 mbar of Ar (grade 5.0) at 1450~{\DEGC} and an Ar flow of 0.1 slm for 15 minutes. \cite{ostler2010a} H-intercalation was achieved by annealing the samples for 75 minutes in 930 mbar hydrogen (grade 8.0) at 550~{\DEGC} and a flow rate of 0.9 slm. Both, N doped 6H-SiC(0001) wafers with a doping concentration of $1\times 10^{18}$~cm$^{-3}$ from SiCrystal as well as semi-insulating 6H-SiC(0001) wafers from II-VI Inc. were used in this study without any noticeable difference in the results. The latter samples were used for electronic transport measurements.

First, we use FTIR-ATR spectroscopy to study the Si-H bonds as shown in fig. \ref{fig:spectrum}(a) for two different polarizations. In p-polarization a sharp absorption line due to the Si-H stretch mode is seen at 2131~cm$^{-1}$, which unambiguously proves the hydrogenation of the SiC(0001) surface underneath QFMLG. The small line width indicates a high degree of order. Position and width of this signal fit very well to observations made on bare H-terminated 6H-SiC(0001) surfaces (2128-2133.5~cm$^{-1}$).\cite{sieber2001a}  The absence of a signal in s-polarization indicates that the Si-H bonds are perpendicular to the surface. Studies of H-terminated Si(111) overgrown with Al$_2$O$_3$ by atomic layer deposition showed that in this case the Si-H stretch mode is redshifted by about 24~cm$^{-1}$ and strongly broadened ($FWHM \approx 64$~cm$^{-1}$) compared to the uncovered surface.\cite{gao2006a} This was taken as evidence for a strong interaction between the Si-H entities and the Al$_2$O$_3$ layer, probably via Si-H$\cdots$Al-O bridges between partially negative hydrogen and partially positive aluminum atoms. Apparently there is no such strong interaction between the Si-H configurations and the graphene layer ontop. This fits well to the notion of quasi-free-standing graphene.

Next we study QFMLG by Raman spectroscopy which is particularly sensitive to the nature of the in-plane bonds. Fig. \ref{fig:spectrum}(b) depicts a typical Raman spectrum of QFMLG showing three lines: the 2D line at 2665~cm$^{-1}$, the G line at 1581~cm$^{-1}$, and the D line at 1336~cm$^{-1}$.
The D line shows the presence of short range defects. The rather small $I(D)/I(G)$ ratio suggests that their density is not too large. The narrow 2D line with a full width at half maximum of 24~cm$^{-1}$ is very well described by a single Lorentzian, which is consistent with single layer graphene. \cite{ferrari2006a,graf2007a} In contrast to MLG, \cite{emtsev2009a,roehrl2008a} where we observed a blue shift of the 2D line indicative of compressive strain, the position of the 2D line of QFMLG is shifted by $\Delta_{2D,\text{strain}} = -14~\mathrm{cm}^{-1}$ to lower wavenumbers compared to exfoliated graphene (2679~cm$^{-1}$).\cite{graf2007a} Note that the effect of doping on the 2D line position is negligible for the carrier concentrations observed in our samples. \cite{das2008a} Using the expression\cite{suppl} $\Delta a/a_0 = -(\Delta_{2D,\text{strain}} / 11.3~\mathrm{cm}^{-1}) \times 10^{-3}$  we calculate a tensile strain of $1.2 \times 10^{-3}$. Compared to the 2D line, the strain induced shift of the G line is reduced by a factor of $\Delta_{2D,\text{strain}}/\Delta_{G,\text{strain}}=2.1$. \cite{suppl} Correcting the observed position of the G line for a downward shift of $\Delta_{G,\text{strain}} = -7$~cm$^{-1}$ caused by strain and assuming the position of undoped and unstrained graphene \cite{das2008a} to be around 1583~cm$^{-1}$ we calculate a charge induced shift of the G line of $\Delta_{G,\text{charge}} = 5~\mathrm{cm}^{-1}$ to higher wavenumbers. This is consistent with a hole doping of $p=4.0 \times 10^{12}$~cm$^{-2}$. \cite{das2008a} The carrier concentration estimated in this way agrees with the observed intensity ratio $I(2D)/I(G)$, which amounts to 1.45 \cite{das2008a}, and with the values determined by Hall effect. While the compressive strain in MLG can be explained by the different thermal expansion of graphene and SiC,\cite{roehrl2008a} the origin of the tensile strain in QFMLG requires further investigation. However, the strain induced by the graphene/substrate interaction in MLG is released in QFMLG which supports the attribute "quasi-free-standing".

In order to characterize the coupling of charge carriers in QFMLG to the substrate we carried out Hall measurements, giving independent values for charge carrier density and mobility. Our first experiments\cite{speck2010a} on QFMLG have resulted in $p=6.0 \times 10^{12}$~cm$^{-2}$ and $\mu = 1,250$~{\MOB} at RT. Recent progress in the preparation results in mobility values of around 3,100~{\MOB} at RT as shown in fig. \ref{fig:transport}(a). The QFMLG samples are hole doped with $p=5.0-6.5\times 10^{12}$~cm$^{-2}$ as displayed in the histogram in fig. \ref{fig:transport}(b). Of special interest is the $T$ dependence of $\mu$ in QFMLG, which is compared to that of MLG \cite{jobst2010a} in fig. \ref{fig:transport}(c). While $\mu$ drops by more than 50\% when $T$ is increased from 25 K to 300 K for MLG, the change in $\mu$ is only 10\% for QFMLG in the same temperature range. Note that for both samples the carrier concentration is basically constant in the whole temperature region. The observed improvement of the temperature dependence points towards a different strength of the interaction between the graphene layer and the substrate. 

Remote phonon scattering has been suggested as a major contribution to the $T$ dependence of $\mu$. \cite{chen2008a,morozov2008a,fratini2008a,zou2010a} Within this scheme, different phonon modes must be taken into account for MLG and QFMLG. Assuming a single phonon mode coupled to the charge carriers in graphene (similar to ref. \cite{zou2010a}), we find phonon energies of 18~meV and 58~meV and coupling strengths of 510~$\Omega$ and 260~$\Omega$ for MLG and QFMLG respectively. The data cannot be explained using different coupling strengths while keeping the same phonon energy. It is not surprising that the substantially differing layer underneath the graphene yields both different coupling strength and different phonon modes.  However, further work is necessary to pinpoint the responsible phonon modes. Hall measurements indicate a substantial charging of the graphene sheet (opposite in sign compared to MLG). This is clearly different from the charge neutral free graphene. The mobility however with its weak temperature dependence justifies the name "quasi-free-standing".

In conlusion we have studied QFMLG on SiC(0001) prepared by conversion of the BL through SiC surface hydrogenation, which was demonstrated  by FTIR-ATR spectroscopy. Raman spectra confirmed that the BL is converted to QFMLG. A negligible interaction between QFMLG and the substrate surface is evident from sharp Si-H stretch mode signals and the lack of compressive strain usually observed in MLG. The QFMLG layers are hole doped and mobilities of 3,100~{\MOB} at RT have been achieved. In contrast to MLG only a small temperature dependence of the mobility is observed. Hence, QFMLG is efficiently decoupled from the substrate and provides an interesting and tunable material for the development of electronic devices. \\

This work was supported by the DFG under contract SE 1087/5-1, SE 1087/9-1, and WE 4542/5-1, by the DFG Priority Program "Graphene", and by the ESF through the program EuroGRAPHENE.


\begin{thebibliography}{3}


\bibitem{berger2006a}
C. Berger, Z.~M. Song, X.~B. Li, X.~S. Wu, N. Brown, C. Naud, D. Mayou, T.~B. Li, J. Hass, A.~N. Marchenkov, E.~H. Conrad, P.~N. First, and W.~A. de Heer, Science \textbf{312}, 1191 (2006)

\bibitem{emtsev2009a}
K.~V. Emtsev, A. Bostwick, K. Horn, J. Jobst, G.~L. Kellogg, L. Ley, J.~L. Mcchesney, T. Ohta, S.~A. Reshanov, J. R\"ohrl, E. Rotenberg, A.~K. Schmid, D. Waldmann, H.~B. Weber, Th. Seyller, Nat. Mater \textbf{8}, 203 (2009)

\bibitem{first2010a}
P.~N. First, W. A. de Heer, Th. Seyller, C. Berger, J. A. Stroscio, J.-S. Moon, MRS Bull. \textbf{35}, 296 (2010)

\bibitem{lin2010a}
Y.-M. Lin, C. Dimitrakopoulos, K. A. Jenkins, D. B. Farmer, H.-Y. Chiu, A. Grill, P. Avouris, Science \textbf{327}, 662 (2010)

\bibitem{tzalenchuk2010a}
A. Tzalenchuk, S. Lara-Avila, A. Kalaboukhov, S. Paolillo, M. Syvajarvi, R. Yakimova, O. Kazakova, T. Janssen, V. Fal'ko, S. Kubatkin, Nat. Nano. \textbf{5}, 186 (2010)

\bibitem{emtsev2008a}
K. V. Emtsev, F. Speck, Th. Seyller, L. Ley, J. D. Riley, Phys. Rev. B. \textbf{77}, 155303 (2008)

\bibitem{bostwick2007a}
A. Bostwick, T. Ohta, Th. Seyller, K. Horn, E. Rotenberg, Nat. Phys. \textbf{3}, 36 (2007)

\bibitem{roehrl2008a}
J. R\"ohrl, M. Hundhausen, K. V. Emtsev, Th. Seyller, R. Graupner, L. Ley, Appl. Phys. Lett. \textbf{92}, 201918 (2008)

\bibitem{jobst2010a}
J. Jobst,  D. Waldmann, F. Speck, R. Hirner, D. K. Maude, Th. Seyller, H. B. Weber, Phys. Rev. B \textbf{81}, 195434 (2010)

\bibitem{shen2009a}
T. Shen, J. J. Gu, M. Xu, Y. Q. Wu, M. L. Bolen, M. A. Capano, L. W. Engel, P. D. Ye, Appl. Phys. Lett. \textbf{95}, 172105 (2009)

\bibitem{kopylov2010a}
S. Kopylov,  A. Tzalenchuk, S. Kubatkin, V. I. Fal'ko, Appl. Phys. Lett. \textbf{97}, 112109 (2010)

\bibitem{novoselov2005a}
K. S. Novoselov, A. K. Geim, S. V. Morozov, D. Jiang, M. I. Katsnelson, I. V. Grigorieva, S. V. Dubonos, A. A. Firsov, Nature \textbf{438}, 197 (2005)

\bibitem{zhang2005a}
Y. Zhang, Y.-W. Tan, H. L. Stormer, P. Kim, Nature \textbf{438}, 201 (2005)

\bibitem{weingart2010a}
S. Weingart, C. Bock, U. Kunze, F. Speck, Th. Seyller, L. Ley, Physica E \textbf{42}, 687 (2010)

\bibitem{riedl2010a}
C. Riedl, C. Coletti, T. Iwasaki, A. A. Zakharov, U. Starke, Phys. Rev. Lett. \textbf{103}, 246804 (2010)

\bibitem{sieber2003a}
N. Sieber, Th. Seyller, L. Ley, D. James, J. D. Riley, R. C. G. Leckey, Phys. Rev. B \textbf{67}, 205304 (2003)

\bibitem{ostler2010a}
M. Ostler, F. Speck, M. Gick, Th. Seyller, Phys. Stat. Sol. B \textbf{247}, 2924 (2010)

\bibitem{sieber2001a}
N. Sieber, B. F. Mantel, Th. Seyller, J. Ristein, L. Ley, T. Heller, D. R. Batchelor, D. Schmeisser, Appl. Phys. Lett. \textbf{78}, 1217 (2001)

\bibitem{gao2006a}
K. Y. Gao, F. Speck, K. Emtsev, Th. Seyller, L. Ley, M. Oswald, W. Hansch, Phys. Stat. Sol. A \textbf{203}, 2194 (2006)

\bibitem{ferrari2006a}
A. C. Ferrari, J. C. Meyer, V. Scardaci, C. Casiraghi, M. Lazzeri, F. Mauri, S. Piscanec, D. Jiang, K. S. Novoselov, S. Roth, A. K. Geim, Phys. Rev. Lett. \textbf{97}, 187401 (2006)

\bibitem{graf2007a}
D. Graf, F. Molitor, K. Ensslin, C. Stampfer, A. Jungen, C. Hierold, L. Wirtz, Nano Lett. \textbf{7}, 238 (2007)

\bibitem{das2008a}
A. Das, S. Pisana, B. Chakraborty, S. Piscanec, S. K. Saha, U. V. Waghmare, K. S. Novoselov, H. R. Krishnamurty, A. K. Geim, A. C. Ferrari, A. K. Sood, Nat. Nanotechnol. \textbf{3}, 210 (2008)

\bibitem{suppl}
See supplementary material at [URL will be inserted by AIP].

\bibitem{speck2010a}
F. Speck, M. Ostler, J. R\"ohrl, J. Jobst, D. Waldmann, M. Hundhausen, L. Ley, H. B. Weber, and Th. Seyller, Mater. Sci. Forum \textbf{645-648}, 629 (2010)

\bibitem{chen2008a}
J. Chen, C. Jang, S. Xiao, M. Ishigami, M. Fuhrer, Nat. Nano. \textbf{3}, 206 (2008)

\bibitem{morozov2008a}
S. V. Morozov, K. S. Novoselov, M. I. Katsnelson, F. Schedin, D. C. Elias, J. A. Jaszczak, A. K. Geim, Phys. Rev. Lett. \textbf{100}, 016602 (2008)

\bibitem{fratini2008a}
S. Fratini and F. Guinea, Phys. Rev. B \textbf{77}, 195415 (2008)

\bibitem{zou2010a}
K. Zou, X. Hong, D. Keefer, J. Zhu, Phys. Rev. Lett. \textbf{105}, 126601 (2010)


\end{thebibliography}
\end{document}